\documentclass[sigconf]{acmart}
\usepackage{booktabs} 

\makeatletter
\renewcommand\@formatdoi[1]{\ignorespaces}
\renewcommand\footnotetextcopyrightpermission[1]{}
\makeatother
\setcopyright{none}

\PassOptionsToPackage{bookmarks=false}{hyperref}
\usepackage{color}
\usepackage{algorithm}
\usepackage{algpseudocode}
\usepackage{pseudocode}
\usepackage{graphicx}
\usepackage{amsmath}
\usepackage{csquotes}
\usepackage{multirow}
\usepackage{tikz}
\usepackage{subcaption}

\def\edit{}

\newcommand{\sys}{DeepFense}

\graphicspath{{figures/}} 

\begin{document}
\copyrightyear{2018} 
\acmYear{2018} 
\setcopyright{acmcopyright}
\acmConference[ICCAD '18]{IEEE/ACM INTERNATIONAL CONFERENCE ON COMPUTER-AIDED DESIGN}{November 5--8, 2018}{San Diego, CA, USA}
\acmBooktitle{IEEE/ACM INTERNATIONAL CONFERENCE ON COMPUTER-AIDED DESIGN (ICCAD '18), November 5--8, 2018, San Diego, CA, USA}
\acmPrice{15.00}
\acmDOI{10.1145/3240765.3240791}
\acmISBN{978-1-4503-5950-4/18/11}

\title{DeepFense: Online Accelerated Defense Against \\ Adversarial Deep Learning}
\author{
    \Large Bita Darvish Rouhani, Mohammad Samragh, Mojan Javaheripi, Tara Javidi, and Farinaz Koushanfar\\
    University of California San Diego\\
    bita@ucsd.edu, msamragh@ucsd.edu, mojan@ucsd.edu, tjavidi@ucsd.edu, farinaz@ucsd.edu
} 


\begin{abstract}
\edit{Recent advances in adversarial Deep Learning (DL) have opened up a largely unexplored surface for malicious attacks jeopardizing the integrity of autonomous DL systems.} With the wide-spread usage of DL in critical and time-sensitive applications, including unmanned vehicles, drones, and video surveillance systems, online detection of malicious inputs is of utmost importance. We propose \sys{}, the \textit{first end-to-end automated} framework that simultaneously enables efficient and safe execution of DL models. \edit{\sys{} formalizes the goal of thwarting adversarial attacks as an optimization problem that minimizes the rarely observed regions in the latent feature space spanned by a DL network. To solve the aforementioned minimization problem, a set of complementary but disjoint modular redundancies are trained to validate the legitimacy of the input samples in parallel with the victim DL model.} \sys{} leverages hardware/software/algorithm co-design and customized acceleration to achieve just-in-time performance in resource-constrained settings. \edit{The proposed countermeasure is unsupervised, meaning that no adversarial sample is leveraged to train modular redundancies.} We further provide an accompanying API to reduce the non-recurring engineering cost and ensure automated adaptation to various platforms. Extensive evaluations on FPGAs and GPUs demonstrate up to two orders of magnitude performance improvement while enabling online adversarial sample detection. 
\end{abstract}
\keywords{Adversarial Attacks, Deep Learning, Model Reliability, FPGA Acceleration, Real-time Computing}
\maketitle

\section{Introduction}
Deep Neural Networks (DNNs) have enabled a transformative shift in various scientific fields ranging from natural language processing and computer vision to health-care and intelligent transportation~\cite{mcdaniel2016machine, deng2014deep, knorr2015paypal}. Although DNNs demonstrate superb accuracy in controlled settings, it has been shown that they are particularly vulnerable to adversarial samples: carefully crafted input instances which lead machine learning algorithms into misclassifying while the input changes are imperceptible to a naked eye.

In response to the various adversarial attack methodologies proposed in the literature (e.g.,~\cite{carlini2017towards,goodfellow2014explaining,kurakin2016adversarial,moosavi2016deepfool}), several research attempts have been made to design DL strategies that are more robust in the face of adversarial examples. \edit{The existing countermeasures, however, encounter (at least) two sets of limitations: (i) Although the prior-art methods have reported promising results in addressing adversarial attacks in black-box settings~\cite{meng2017magnet, zantedeschi2017efficient, shen2017ape}, their performance has been shown to significantly drop in white-box scenarios where the adversary has the full knowledge of the defense mechanism~\cite{carlini2017magnet}. (ii)} None of the prior works have provided an automated hardware-accelerated system for online defense against adversarial inputs. Due to the wide-scale adoption of DL in sensitive autonomous scenarios, it is crucial to equip all such models with a defense mechanism against the aforementioned adversaries. 

We propose \sys{}, the first end-to-end hardware-accelerated framework that enables robust and just-in-time defense against adversarial attacks on DL models. \edit{Our key observation is that the vulnerability of DNNs to adversarial samples originates from the existence of rarely-explored sub-spaces spanned by the activation maps in each (hidden) layer. This phenomenon is particularly caused by (i) the high dimensionality of activation maps and (ii) the limited amount of labeled data to fully traverse/learn the underlying space.} To characterize and thwart potential adversarial sub-spaces, we propose a new method called \textit{Modular Robust Redundancy} (MRR). \edit{MRR is robust against the state-of-the-art adaptive white-box attacks in which the adversary knows everything about the victim model and its defenders.}

\edit{Each modular redundancy characterizes the explored subspace in a given layer by learning the Probability Density Function (PDF) of typical data points and marking the complement regions as rarely-explored/risky.} Once such characterization is obtained, the checkpointing modules evaluate the input sample in parallel with the victim model and raise alarm flags for data points that lie within the risky regions. The MRRs are trained in unsupervised settings meaning that the training dataset is merely composed of typical benign samples. This, in turn, ensures resiliency against potential new attacks. Our unsupervised countermeasure impacts neither the training complexity nor the final accuracy of the victim DNN. 

\begin{figure*}
\centering
\includegraphics[width=0.75\textwidth]{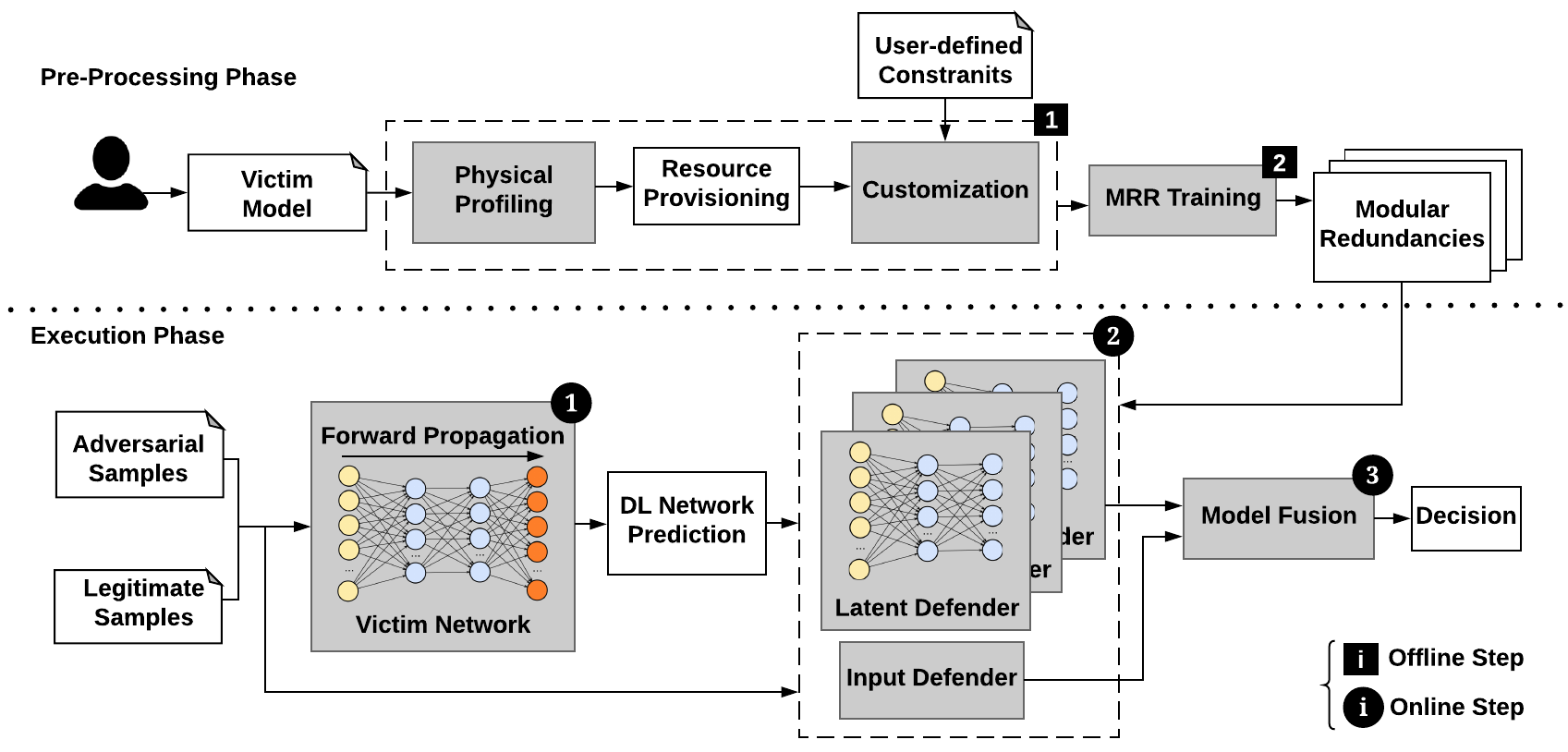}{\centering}
\vspace{-1.2em}
\caption{ \small Global flow of the \sys{} framework. \sys{} takes as input the high-level description of a DL model together with the proposed defender topologies. Based on the user-provided constraints, \sys{} outputs the best defense layout to ensure maximum throughput and power efficiency, customized for the resource-constrained target hardware platform.}
\label{fig:global_flow}
\vspace{-1.1em}
\end{figure*}

\sys{} is devised based on a hardware/software/algorithm co-design approach to enable safe DL while customizing system performance in terms of latency, energy consumption, and/or memory footprint with respect to the underlying resource provisioning. There is a trade-off between system performance and robustness against adversarial attacks that is determined by the number of modular redundancies. \sys{} provides an automated tool to adaptively maximize the robustness of the defense model while adhering to the user-defined and/or hardware-specific constraints. We chose FPGAs to provide fine-grained parallelism and just-in-time response by our defender modules. The customized data path for memory access and network schemes on FPGA, in turn, helps to improve the overall system energy efficiency.

Although several hardware-accelerated tools for DL execution have been proposed in the literature, e.g.~\cite{zhang2015optimizing, chen2014diannao,sharma2016dnnweaver,samragh2017customizing,rouhani2017deep3}, none of them have been particularly optimized for in-time defense against adversarial inputs. For instance, defenders require a custom layer to characterize and compare each incoming data sample against the PDF of legitimate data. These types of custom layers are atypical to conventional DNNs and have not been addressed in prior works. In summary, the contributions of this paper are as follows: 

\begin{itemize}
    \item Proposing \sys{}, the first hardware/software/algorithm co-design that empowers online defense against adversarial samples for DNNs. \edit{\sys{} methodology is unsupervised and robust against the most challenging attack scenario in real-world applications (white-box attacks).}  
    \item Devising an automated customization tool to adaptively maximize DL robustness against adversarial samples while complying with the underlying hardware resource constraints in terms of run-time, energy, and memory footprint.
    \item Providing the first implementation of custom streaming-based DL defense using FPGAs. \sys{} leverages dictionary learning and probability density functions to statistically detect abnormalities in the inputted data samples. 
    \item Performing extensive proof-of-concept evaluations on common DL benchmarks against the state-of-the-art adversarial attacks reported to-date. \edit{Thorough performance comparison on various hardware platforms including embedded CPUs, GPUs, and FPGAs corroborates \sys{}'s efficiency.}
\end{itemize}

\vspace{-0.5em}
\section{D\MakeLowercase{eep}F\MakeLowercase{ense} Global Flow}
Figure~\ref{fig:global_flow} illustrates the global flow of \sys{} framework. We consider a system consisting of a single classifier (a.k.a., victim model) and a set of defender modules aiming to detect adversarial samples. \edit{\sys{} consists of two main phases to characterize and thwart adversarial attacks: (i) offline pre-processing phase to train defender modules, and (ii) online execution phase in which the legitimacy of each incoming input data is validated on the fly. The one-time pre-processing phase is performed in software while the recurrent execution phase is accelerated using FPGA.}

\vspace{0.2em}
\noindent \textbf{\large Pre-processing phase.} This phase consists of two tasks.  

\noindent {\tikz\draw[black,fill=black] (-1em,-1em) rectangle (-0.2em,-0.2em) node[pos=.5, white] {1};} \textbf{Resource Profiling and Design Customization.} \edit{There is a trade-off between execution run-time and system reliability in terms of successful adversarial detection rate.} \sys{} uses physical profiling to estimate resource utilization for the victim model as well as the defender modules. \edit{The output of physical profiling along with a set of user-defined constraints (e.g., real-time requirements) is then fed into the design customization unit to determine the viable number of defenders and their appropriate locations based on the sensitivity of DNN layers (Section~\ref{customization}).} The customization unit analyzes the trade-off between model reliability, resource limitation, and throughput to decide the best combination of defenders suitable to the task and customized for the target hardware. 

\vspace{0.1em}
\noindent {\tikz\draw[black,fill=black] (-1em,-1em) rectangle (-0.2em,-0.2em) node[pos=.5, white] {2};} \textbf{Training Modular Redundancies.} \edit{\sys{} trains a set of redundancy modules (checkpoints) to isolate potential adversarial sub-spaces.}  The redundancy modules can be categorized into two classes, namely the \textit{Input Defenders} (Section~\ref{input_defender}) and the \textit{Latent (Intermediate) Defenders} (Section~\ref{latent_defender}). \edit{Each defender targets a particular layer in the victim model and is trained with the goal of separating data manifolds and characterizing the underlying PDF by careful realignment of legitimate data within each class.}

\vspace{0.1em}
\noindent \textbf{\large Execution phase.} \edit{Once the redundancy modules are trained and customized per hardware and/or user-defined physical constraints, the underlying DL model is ready to be deployed for online execution. \sys{} performs three tasks for the execution phase.}

\begin{table*}
\centering
\caption{Motivational example: We compare the MRR methodology against prior-art works including Magnet~\cite{meng2017magnet}, Efficient Defenses Against Adversarial Attacks~\cite{zantedeschi2017efficient}, and APE-GAN~\cite{shen2017ape} in the white-box setting. For each evaluation, the adversarial perturbation ($L_2$ distortion) is normalized to that of the attack without the presence of any defense mechanism.}
\vspace{-0.75em}
\label{tab:white_box_attack}
\resizebox{0.97\textwidth}{!}{
\begin{tabular}{l||cccccc||cccccc||c|c|c}& \multicolumn{12}{c||}{\textbf{MRR Methodology}}                    & \multicolumn{3}{c}{\textbf{Prior-Art Defenses}} \\ \cline{2-16} \cline{2-16}
\textbf{Security Parameter}&\multicolumn{6}{c||}{SP=1\%}&\multicolumn{6}{c||}{SP=5\%}& Magnet&Efficient Defenses&APE-GAN\\ \hline
\textbf{Number of Defenders }  & N=0& N=1& N=2& N=4& N=8& N=16&  N=0& N=1& N=2& N=4& N=8& N=16  &N=16& - & - \\
\textbf{Defense Success} & - & 43\% & 53\% & 64\% & 65\% & 66\% & - & 46\% & 63\% & 69\% & 81\% & 84\%  & 1\% & 0\% & 0\%\\
\textbf{Normalized Distortion ($L_2$)} & 1.00 &1.04 &1.11 &1.12 &1.31 &1.38 &1.00 &1.09 &1.28 &1.28 &1.63& 1.57 &1.37 &1.30 &1.06 \\
\textbf{FP Rate}& -& 2.9\%& 4.4\%& 6.1\%& 7.8\%& 8.4\%& -& 6.9\%& 11.2\%& 16.2\%& 21.9\%& 27.6\%& -& -& -            
\end{tabular}
}
\vspace{-0.5em}
\end{table*}

\vspace{0.1em}
\noindent {\tikz\draw[black,fill=black] (0,0) circle (1.1ex) node[white] {1};} \textbf{Forward Propagation.} \edit{The predicted class for each incoming sample is acquired through forward propagation in the victim DNN. The predicted output is then fed to the defenders for validation.}

\vspace{0.1em}
\noindent {\tikz\draw[black,fill=black] (0,0) circle (1.1ex) node[white] {2};} \textbf{Validation.} \edit{\sys{} leverages the checkpoints learned in the pre-processing phase to validate the legitimacy of the input data and the associated label determined in the forward propagation step. In particular, samples that do not lie in the \textit{user-defined} probability interval which we refer to as the \textit{Security Parameter (SP)} are discarded as suspicious samples. SP is a constant number in the range of $[0-100]$ which determines the hardness of adversarial detectors. For applications with excessive security requirements, a high SP value assures full detection of adversarial samples.}

\vspace{0.1em}
\noindent {\tikz\draw[black,fill=black] (0,0) circle (1.1ex) node[white] {3};} \textbf{Model Fusion.} \edit{The outputs of the redundancy modules are finally aggregated to compute the legitimacy probability of the input data and its associated inference label (Section~\ref{sec:modelfusion}).} 

\vspace{0.5em}
\noindent \textbf{Attack Model.} We consider the \textit{adaptive white-box} threat model as the most powerful attacker that can appear in real-world DL applications. In this scenario, we assume the attacker knows everything about the victim model including the learning algorithm, model topology, and parameters. With the presence of \sys{} parallel defenders, the adversary is required to mislead \emph{all} defenders to succeed in forging an adversarial sample as a legitimate input.

\section{D\MakeLowercase{eep}F\MakeLowercase{ense} Methodology}\label{framework}
\edit{\sys{} trains a number of modular redundancies to characterize the data density distribution in the space spanned by the victim model. In this section, we first provide a motivational example for MRR methodology. We then elaborate on the MRR modules that checkpoint the intermediate DL layers (latent defenders) and input space (input defenders). Lastly, we discuss the model fusion to aggregate the MRR outputs and derive the final decision.}


\subsection{Motivational Example}
\edit{The rationale behind our MRR methodology is not only to thwart adversarial attacks in black-box settings (where the adversary is not aware of the defense mechanism), but also to boost the reliability of the model prediction in presence of adaptive white-box attacks. Table~\ref{tab:white_box_attack} compares the success rate of the adaptive white-box Carlini\&WagnerL2 attack~\cite{carlini2017magnet} against MRR methodology with the prior-art countermeasures on MNIST benchmark.\footnote{We used the open-source library https://github.com/carlini/MagNet to implement the Carlini\&WagnerL2 adaptive attack.} We define the \textit{False Positive (FP)} rate as the ratio of legitimate test samples that are mistaken for adversarial samples by \sys{}. The \textit{True Positive (TP)} rate is defined as the ratio of adversarial samples detected by \sys{}. As shown, increasing the number of MRR modules not only decreases the attack success rate but also yields a higher perturbation in the generated adversarial samples. The superior performance of \sys{} is associated with learning the distribution of legitimate samples as opposed to prior works which target altering the decision boundaries.}

\vspace{-0.5em}
\subsection{Latent Defenders} \label{latent_defender}
\edit{Each latent defender module placed at the $n^{th}$ layer of the victim model is a neural network architecturally identical to the victim. This homogeneity of topology enables the defenders to suitably address the vulnerabilities of the victim network. We consider a Gaussian Mixture Model (GMM) as the prior probability to characterize the data distribution at each checkpoint location. We emphasize that our proposed approach is rather generic and is not restricted to the GMM. The GMM distribution can be replaced with any other prior depending on the application.}

\vspace{0.5em}
\noindent \textbf{\large Training a single latent defender.} \edit{To effectively characterize the explored sub-space as a GMM distribution, one is required to minimize the entanglement between every two Gaussian distributions (corresponding to every two different classes) while decreasing the inner-class diversity. There are three main steps that shall be performed to train one latent defender module.}

\vspace{0.2em}
\noindent \textbf{Step 1.} \edit{Replicating the victim neural network and all its parameters. An $L_2$ normalization layer is inserted in the desired checkpoint location. The normalization layer maps the latent features (activations), $f(x)$, into the Euclidean space such that the acquired activation maps are bounded to a hyper-sphere, i.e.,  $\|f(x)\|_2 = 1$. This normalization is crucial as it partially removes the effect of over-fitting to particular data samples that are highly correlated with the underlying DL parameters.}

\vspace{0.2em}
\noindent \textbf{Step 2.} \edit{Fine-tuning the replicated neural network to enforce disentanglement of data features (at a particular checkpoint location) and characterize the PDF of explored sub-spaces. To do so, we optimize the defender module by adding the following loss function to the conventional cross entropy loss: 
\begin{equation} \label{eq:opt}
\scalebox{0.9}{
$\gamma~[~\underbrace{\|C^{y^*} - f(x)\|_2^2}_{{loss_1}}
 ~-~ \underbrace{\Sigma_{i\ne y^*} \|C^i - f(x)\|_2^2}_{loss_2}  ~+~ \underbrace{\Sigma_{i} (\|C^i\|_2 - 1)^2}_{loss_3}~].$}
\end{equation}
Here, $\gamma$ is a trade-off parameter that specifies the contribution of the additive loss term, $f(x)$ is the corresponding feature vector of input sample $x$ at the checkpoint location, $y^*$ is the ground-truth label, and $C^i$ denotes the center corresponding to class $i$. The center values $C^i$ and intermediate feature vectors $f(x)$ are trainable variables that are learned by fine-tuning the defender module. In our experiments, we set the parameter $\gamma$ to $0.01$ and retrain the defender with the same optimizer used for training the victim DNN. The learning rate is set to $0.1$ of that of the victim model as the model is already in a relatively good local minimum.}

\edit{The first term ($loss_1$) in Eq. (\ref{eq:opt}) aims to condense latent data features $f(x)$ that belong to the same class. Reducing the inner-class diversity, in turn, yields a sharper Gaussian distribution per class. The second term ($loss_2$) intends to increase the intra-class distance between different categories and promote separability. If the loss function consists solely of the first two terms in Eq. (\ref{eq:opt}), the pertinent model may diverge by pushing the centers to $C^i\mapsto \pm\infty$. We add the term, $loss_3$, to ensure that the pertinent centers lie on a unit hyper-sphere and avoid divergence.}

\vspace{0.2em}
\noindent \textbf{Step 3.} \edit{After applying Step 2, the latent data features are mapped to discrete GMMs. Each GMM is defined by the first order (mean) and second order statistics (covariance) of the legitimate activations. Using the obtained distributions, \sys{} profiles the percentage of benign samples lying within different $L_2$ radius of each GMM center. We leverage a security parameter in the range of $[0-100]$ to divide the underlying space into the sub-space where the legitimate data lives and its complementary adversarial sub-space. The acquired percentile profiling is employed to translate the user-defined SP into an $L_2$ threshold which is later used to detect malicious samples during the online execution phase.}

\begin{figure}[ht!]
\centering
\includegraphics[width=0.23\textwidth]{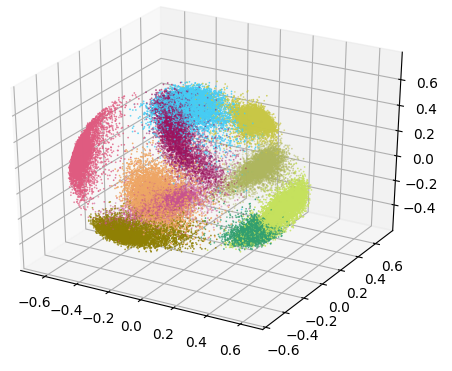}
\includegraphics[width=0.23\textwidth]{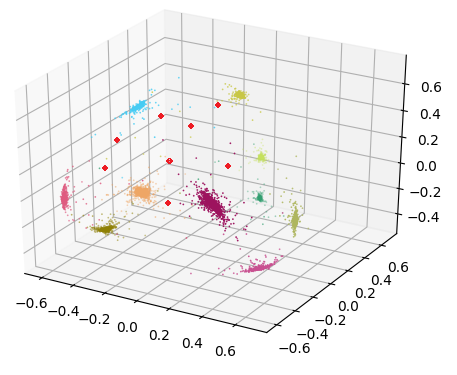}
\caption{\label{fig:step3} Example feature samples in the second-to-last layer of LeNet3 model trained for classifying MNIST data before (left figure) and after (right figure) data realignment performed in Step 2. The majority of adversarial samples (the red dot points) reside in the low density regions.}
\end{figure}

\edit{Figure~\ref{fig:step3} illustrates the activation maps in the second-to-last layer of a LeNet3 network trained for classifying MNIST data before and after data realignment. As shown, the majority of adversarial samples reside in the rarely-explored regions that can be effectively detected by our latent defenders.}

\vspace{0.5em}
\noindent \textbf{\large Training multiple negatively correlated defenders.} \edit{The reliability of MRR defense can be increased by training multiple defenders per layer that are negatively correlated, as opposed to using only one latent defender. Consider a defender module that maps a legitimate input $x$ to the feature vector $f(x)$, where $f(x)$ is close (in terms of Euclidean distance) to the corresponding center $C^i$. An adversary trying to mislead this defender would generate a perturbed input $x+\eta$ such that $f(x+\eta)$ is far from $C^i$ and close to another target center $C^{j}$. In other words, the adversary would like to increase the $loss_1$ term in Eq.~(\ref{eq:opt}). To mitigate such adaptive attacks, we propose to train a Markov chain of defenders.}

\edit{To build the corresponding Markov chain of latent defenders, we start off by training a single defender module as described earlier in this section. Next, we generate a new set of training data that can enforce negative correlations between the current defender module and the next defender. In particular, the $n^{th}$ defender of this chain takes an input data $x$, generates a perturbation $\eta$, and feeds $clip(x+\eta)$ to the $(n+1)^{th}$ defender. The $clip(\cdot)$ operation simply clips the input sample in a valid range of numerical values, e.g., between 0 and 1. The perturbation $\eta$ is chosen as $\eta=\frac{\partial loss_1}{\partial x}$, where the $loss_1$ term (See Eq.~(\ref{eq:opt})) corresponds to the $n^{th}$ defender. Given this new dataset of perturbed samples, benign data points that deviate from the centers in the $n^{th}$ defender will be close to the corresponding center in the $(n+1)^{th}$ defender. As such, simultaneously deceiving all the defenders requires a higher amount of perturbation.}


\subsection{Input Defender} \label{input_defender}
\edit{One may speculate that an adversary can add a structured noise to a legitimate sample such that the data point is moved from one cluster to the center of the other clusters; thus fooling the latent defender modules. The risk of such an attack approach is significantly reduced by leveraging sparse signal recovery techniques. We use dictionary learning to measure the Peak Signal-to-Noise Ratio (PSNR) of each incoming data and filter out atypical samples in the input space. An input checkpoint is configured in two main steps.}

\vspace{0.2em}
\noindent \textbf{Step 1.} \edit{We learn a separate dictionary for each class by solving:
\begin{equation}
\scalebox{0.85}{
$\begin{aligned}
\underset{D^i}{argmin}~\frac{1}{2}\|Z^i - D^iV^i\|_2^2 + \beta \|V^i\|_1~~~s.t.~~~\|D_k^i\| = 1,~~ 0 \leq k \leq k_{max}.
\end{aligned}\label{eq:dictionary_objective}$}
\vspace{-0.2em}
\end{equation}
Here, $Z^i$ is a matrix whose columns are pixels extracted from different regions of input images belonging to category $i$. For instance, if we consider $8\times 8$ patches of pixels, each column of $Z^i$ would be a vector of $64$ elements. The goal of dictionary learning is to find matrix $D^i$ that best represents the distribution of pixel patches from images belonging to class $i$. We denote the number of columns in $D^i$ by $k_{max}$. For a certain $D^i$, the image patches $Z^i$ are represented with a sparse matrix $V^i$, and $D^iV^i$ is the reconstructed sample. We leverage Least Angle Regression (LAR) to solve Eq. (\ref{eq:dictionary_objective}).}

During the execution phase, the input defender module takes the output of the victim DNN (e.g., predicted class $i$) and uses Orthogonal Matching Pursuit (OMP) \cite{tropp2007signal} to sparsely reconstruct the input with the corresponding dictionary $D^i$. The input sample labeled as class $i$ should be well-reconstructed as $D^iV^*$ with a high PSNR value, where $V^*$ is the optimal solution obtained by OMP.

\vspace{0.2em}
\noindent \textbf{Step 2.} \edit{We profile the PSNR percentiles of legitimate samples within each class and find the corresponding threshold that satisfies the user-defined security parameter. If an incoming sample has a PSNR lower than the threshold (i.e., high perturbation after reconstruction by the corresponding dictionary), it is regarded as malicious data.}  

\begin{figure}[ht!]
\centering
\includegraphics[width=0.2\textwidth]{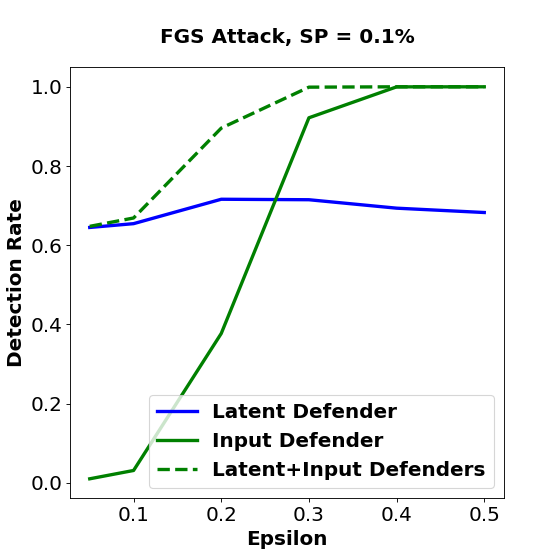}
\includegraphics[width=0.2\textwidth]{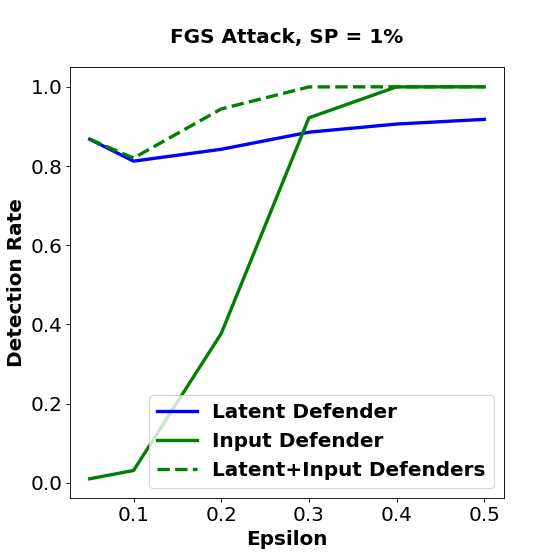}
\caption{\label{fig:errordetect} Adversarial detection rate of the latent and input defender modules as a function of the perturbation level.}
\end{figure}

\edit{Figure~\ref{fig:errordetect} demonstrates the impact of perturbation level ($\epsilon$) on the adversarial detection rate for two different security parameters (cut-off thresholds). In this experiment, we have considered the Fast Gradient Sign (FGS) attack~\cite{goodfellow2014explaining} on LeNet3 MNIST benchmark with a single latent defender inserted at the second-to-last layer. As shown, the use of input dictionaries facilitates detection of adversarial samples with relatively high perturbations.}

\subsection{Model Fusion}{\label{sec:modelfusion}}
\edit{Each defender module in \sys{} framework takes as input a sample $x$ and generates a binary output $d_n\in\{0,1\}$ with value 1 denoting an adversarial sample. This binary decision is based on the user-defined security parameter. To aggregate the binary random variables $\{d_1, \dots, d_N\}$ into a single decision $a$, we compute the probability of the input being adversarial as the following:
\begin{equation}{\label{eq:noor}}
\vspace{-0.3em}
\begin{tabular}{l}
     $P(a=1|\{d_1, d_2, \dots, d_N\})=1-{\displaystyle \prod_{n=1}^{N} (1-P_n)^{d_n}}$, \\
     $P_n=P(a=1|d_n=1)$.
\end{tabular}
\end{equation}
This formulation resembles the well-known noisy-OR terminology used in statistical learning~\cite{diez1993parameter}. In MRR methodology, each defender has a parameter $P_n$ which indicates the likelihood of a sample being adversarial given that the $n^{th}$ defender has labeled it as a malicious sample. If all detectors have $P_n=1$, then the formulation in Eq.~(\ref{eq:noor}) is equivalent to the logical OR between $\{d_1, \dots, d_N\}$.}

\edit{The $P_n$ parameters can be estimated by evaluating the performance of each individual defender. For this purpose, we use a subset of the training data and create adversarial samples with different attack algorithms. If the defender suspects $M_{False}$ samples of the legitimate training data and $M_{True}$ samples of the adversarial data set, the probability $P(a=1|d_n=1)$ is estimated as:
\begin{equation}
P_n=\frac{M_{True}}{M_{False}+M_{True}}.\label{eq:train_noisy_or}
\end{equation}
The output of the noisy-OR model is in the unit interval. \sys{} raises alarm flags for samples with $P(a=1|\{d_1, d_2, \dots, d_n\})\geq 0.5$.}

\section{D\MakeLowercase{eep}F\MakeLowercase{ense} Hardware Acceleration}
\edit{In this section, we first discuss the hardware architecture of latent and input defenders that enables a high throughput and low energy realization of recurrent execution phase. We, then, discuss the resource profiling and automated design customization unit.} 

\subsection{Latent Defenders}{\label{sec:HWLatent}}
During execution, each incoming sample is passed through the latent defender modules that are trained offline (Section~\ref{latent_defender}). The legitimacy probability of each sample is then approximated by measuring the $L_2$ distance with the corresponding GMM center. The latent defenders can be situated in any layer of the victim network, therefore, the extracted feature vector from the DNN can be of high cardinality. High dimensionality of the GMM centers may cause shortage of memory as well as increasing the computational cost and system latency. In order to mitigate the curse of dimensionality, we perform Principal Component Analysis (PCA) on the outputs of the latent defenders before measuring the $L_2$ distance. For the latent defenders in the \sys{} framework, PCA is performed such that more than $99\%$ of the energy is preserved. 

\begin{figure}[ht!]
\centering
\includegraphics[width=0.48\textwidth]{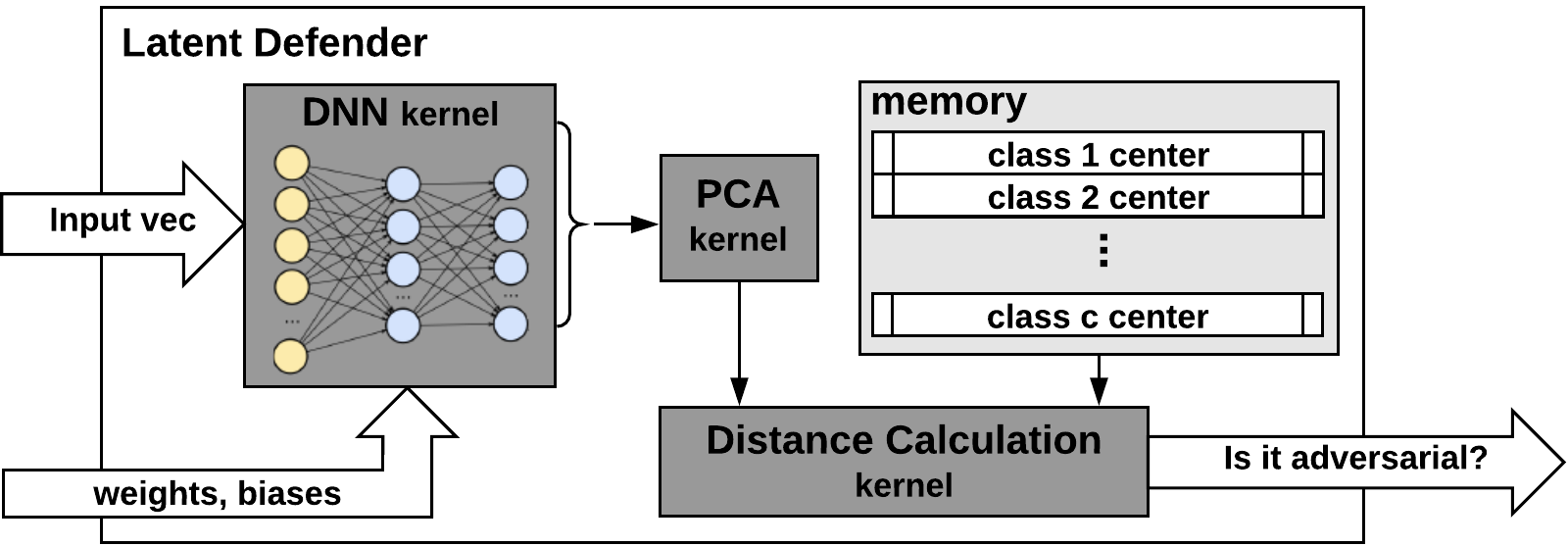}{\centering}
\caption{\small Latent defender structure: The pertinent activations are acquired by propagating the input sample through the defender. PCA is then applied to reduce the dimensionality of the obtained activation. The $L2$ distance with the corresponding GMM center determines the legitimacy of the input.}
\label{fig:latent_defender}
\end{figure}

The most computationally-intensive operation in DNN execution is matrix-matrix multiplication. Recent FPGAs provide hardened DSP units together with the re-configurable logic to offer a high computation capacity. The basic function of a DSP unit is a multiplication and accumulation (MAC). In order to optimize the design and make use of the efficient DSP slices, we took a parallelized approach to convert the DNN layer computations into multiple operations running simultaneously as suggested in ~\cite{sharma2016dnnweaver}. Figure~\ref{fig:latent_defender} illustrates the high-level schematic of a latent defender kernel.

Two levels of parallelism are applied in the implementation of the DNN layers, controlled by parameters $N_{PE}$ and $N_{PU}$ which denote the parallelism level in the input processing and output generation stage, respectively. The aforementioned parameters are static across all layers of the DNN model. In order to achieve maximum throughput, it is essential to fine-tune the parallelism parameters. An increase in the number of parallel computation units will not always result in  better throughput since the dimensionality of the data and divisibility into ready-to-process batches highly affects the efficiency of these parallel units. Section~\ref{customization} provides the details of our optimization method for such parameters.

To minimize the latency of latent defenders, we infuse the PCA kernel into the defender DNNs. Collectively, all transformations from the original input space to the space spanned by principal components can be shown as a vector-matrix multiplication $T=XW_L$ where $W_L$ is a matrix whose columns are Eigenvectors obtained from the legitimate data. The transformation $T=XW_L$ maps a data vector $X$ from an original space of $p$ variables to a new space of $L$ uncorrelated variables. As such, the PCA kernel can be replaced with a \textit{Dense} layer, appended to the defender DNN architecture. 

\subsection{Input Defenders}
\begin{figure}[ht!]
\centering
\includegraphics[width=0.38\textwidth]{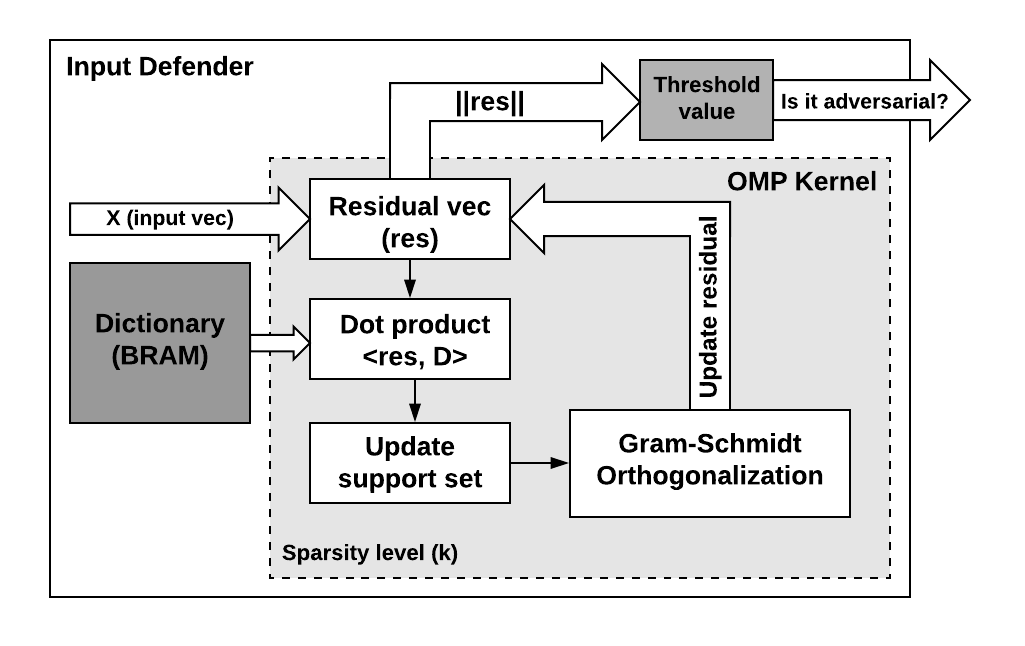}{\centering}
\vspace{-1em}
\caption{\small Input defender structure: The OMP core iteratively reconstructs the input vector by means of a previously learned dictionary. The reconstruction error is used to determine the input legitimacy.}
  \label{fig:input_defender}
\end{figure}

{\label{sec:HWInput}}
Execution of the OMP algorithm is the main computational bottleneck in the input defender modules. The OMP routine requires iterative processing of three main steps: (i) finding the best matching sample in the dictionary matrix D, (ii) Least-Square (LS) optimization, and (iii) residual update. We provide a scalable implementation of the OMP routine on FPGA to enable low-energy and in-time analysis of input data. Figure~\ref{fig:input_defender} illustrates the high-level schematic of an input defender's kernel. Here, the \textit{support set} contains columns of the dictionary matrix that have been chosen so far in the routine. The LS optimization step is performed using Gram-Schmidt orthogonalization technique as suggested in~\cite{rouhani2015ssketch} to reduce the hardware implementation complexity. Note that since the decision of the defender solely depends on the norm of the residual vector, $||res||_2$, there is no need to explicitly compute the sparse vector $V$.

The execution of OMP includes two computationally expensive steps, namely the matrix-vector multiplication and the LS optimization. Each of these steps includes multiple dot product computations. Due to the sequential nature of dot operations, use of pipelining does not improve the throughput. Thereby, we use a tree-based reduction technique to find the final value by adding up the partial results produced by each of the parallel processes. We pipeline and unroll the tree-based reduction function to provide a more efficient solution. Cyclic array partitioning along with loop unrolling is leveraged to ensure maximum throughput and performance.

\subsection{Automated Design Customization}\label{customization}
We provide an automated customization unit that maximizes the robustness of a DNN within the limits of the pertinent resource provisioning. Our automated optimization ensures ease of use and reduces the non-recurring engineering cost. \sys{}'s customization unit takes as input the high-level description of the defenders in Caffe together with the available resources in terms of storage, DSP units, and run-time. It then outputs the best combination of defender modules to ensure maximum robustness against adversarial attacks while adhering to the available resources. We thoroughly examined the performance and resource utilization for different building blocks of a DNN. These blocks include the essential hyper-parameters for instantiating the desired DL model, including but not limited to the number of layers and the corresponding input/output sizes. This enables \sys{} to estimate the upper bound for implementation of a DNN on resource-constrained platforms.

Dictionary matrices leveraged in the input defender as well as the weights and biases of the latent defenders are stored in the on-chip DRAM memory to be accessed during the execution phase. Upon computation, data is moved from the DRAM to Block RAMs (BRAM) which enable faster computations. Our evaluations on various FPGAs show that the main resource bottlenecks are the BRAM capacity and the number of DSP units. As such, \sys{} optimizes the configuration of the defenders with regard to these two constraints. In particular, \sys{} solves the following optimization to find the best configuration for the number of defenders $N_{def}$ and the number of processing units $N_{PU}$ per defender. 
\begin{equation} \label{eq:dict}
\vspace{-1em}
\begin{aligned}
&\underset{N_{PU}, N_{def}}{Maximize}~(DL_{robustness})~~~~~~~s.t.:\\ 
&T^{max}_{def} \leq T_u,~~~~N_{def}\times N_{PU}\times DSP_{PU} \leq R_u,\\
&N_{PU}\times [{max(size({W^i})}) + {max(|X^{i}|+|X^{i+1}|)}]\leq M_u,\\
\end{aligned}
\end{equation}

\vspace{1em}
\noindent where $T_u$, $M_u$, and $R_u$ are user-defined constraints for system latency, BRAM budget, and available DSP resources, respectively. Here, $size(W^i)$ denotes the total number of parameters and $|X^i|$ is the cardinality of the input activation in layer $i$. $DSP_{PU}$ indicates the number of DSP slices used in one processing unit. Variable $T^{max}_{def}$ is the maximum required latency for executing the defender modules. \sys{} considers both sequential and parallel execution of defenders based on the available resource provisioning and size of the victim DNN. Once the optimization is solved for $N_{PU}$, $N_{PE}$ is uniquely determined based on available resources.

The OMP unit in \sys{} incurs a fixed memory footprint and latency for a given application. As such, the optimization of Eq.~(\ref{eq:dict}) does not include this constant overhead. Instead, we exclude this overhead from the user-defined constraints and use the updated upper bounds. In particular, for an OMP kernel, the required computation time can be estimated as $\beta n(kl+k^2)$ where $n$ indicates the number of elements in the input vector, $l$ is the dictionary size, and $k$ represents the sparsity level. $\beta$ is system-dependant and denotes the number of cycles for one floating point operation. The memory footprint for OMP kernel is merely a function of the dictionary size.

Our customization unit is designed such that it maximizes the resource utilization to ensure maximum throughput. \sys{} performs an exhaustive search over the parameter $N_{PU}$ and solves the equations in ~\ref{eq:dict} using the Karush-Kuhn-Tucker (KKT) method to calculate $N_{def}$. The calculated parameters capture the best trade-off between security robustness and throughput. Our optimization outputs the most efficient layout of defender modules as well as the sequential or parallel realization of defenders. This constraint-driven optimization is non-recurring and incurs a negligible overhead.

\section{Experiments}\label{experiments}
We evaluate \sys{} on three different DNN architectures outlined in Table~\ref{tab:architectures}. Each DNN corresponds to one dataset: MNIST, SVHN, and CIFAR-10. We report the robustness of the aforementioned models against four different attacks. The customized defense layout for each network is implemented on two FPGA platforms. A detailed analysis is provided to compare our FPGA implementation with highly-optimized realizations on CPUs and GPUs.

\begin{table}[ht!]
\centering
\caption{Victim DNN architectures. Here, $\mathbf{20C5}$ denotes a convolutional layer with $20$ output channels and $\mathbf{5\times5}$ filters, $\mathbf{MP2}$ indicates a $\mathbf{2\times2}$ max-pooling, $\mathbf{500FC}$ is a fully-connected layer with $\mathbf{500}$ neurons, and $\mathbf{GAP}$ is global average pooling. All hidden layers have a ``Relu'' activation function.}
\vspace{-0.3em}
\label{tab:architectures}
\resizebox{0.95\columnwidth}{!}{%
\begin{tabular}{c||l}
\multicolumn{1}{l||}{\bf{Benchmark}} & \multicolumn{1}{c}{\bf{Architecture}}                                                                                                                        \\ \hline \hline
\bf{MNIST}                          & \begin{tabular}[c]{@{}l@{}}$(input)~1\times 28\times 28-20C5-MP2-50C5-MP2$ \\$-500FC-10FC$\end{tabular}                       \\ \hline
\bf{SVHN}                        & \begin{tabular}[c]{@{}l@{}}$(input)~3\times 32\times 32-20C5-MP2-50C5-MP2$ \\$-1000FC-500FC-10FC$\end{tabular} \\ \hline
\bf{CIFAR-10}                        & \begin{tabular}[c]{@{}l@{}}$(input)~3\times 32\times 32-96C3-96C3-96C3-MP2$\\$-192C3-192C3-192C3-MP2-192C3-192C1$\\$-10C1-GAP-10FC$\end{tabular}                        \\
\end{tabular}
}
\vspace{-1em}
\end{table}

 \begin{figure*}[ht!]
\centering
\includegraphics[width=0.28\textwidth]{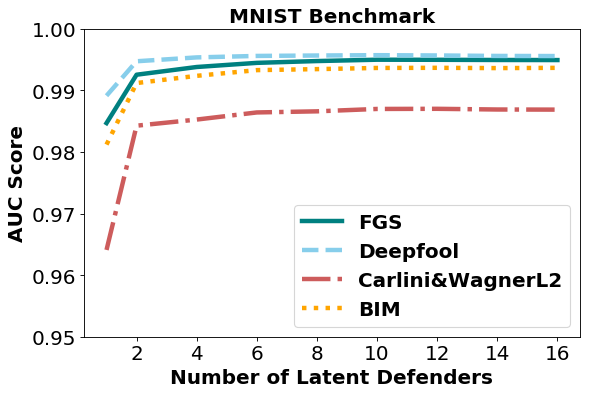}
\includegraphics[width=0.28\textwidth]{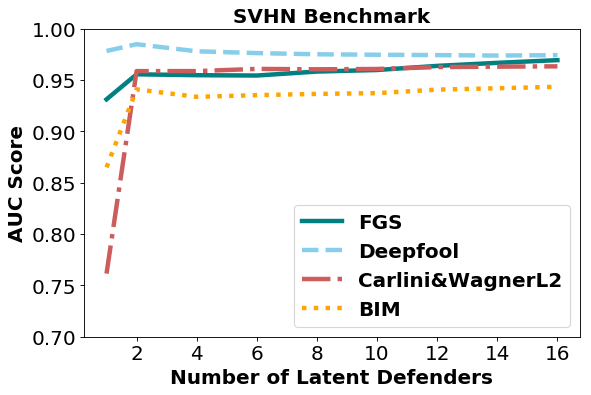}
\includegraphics[width=0.28\textwidth]{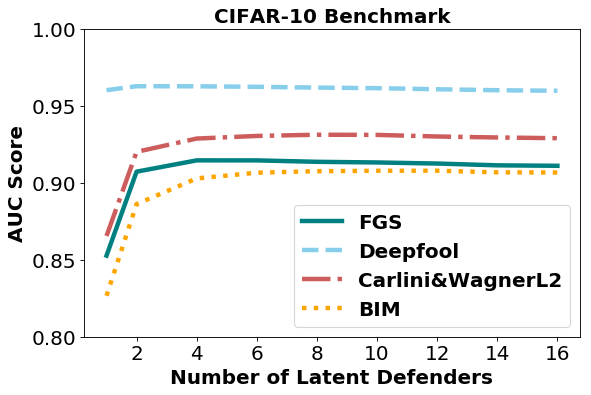}
\caption{\label{fig:auc} AUC score versus the number of defender modules for MNIST, SVHN, and CIFAR-10 datasets.}
\vspace{-1em}
\end{figure*}

\subsection{Attack Analysis and Resiliency}\label{attack_analysis}
\edit{We leverage a wide range of attack methodologies (namely, FGS~\cite{goodfellow2014explaining}, BIM~\cite{kurakin2016adversarial}, CarliniL2~\cite{carlini2017towards}, and Deepfool~\cite{moosavi2016deepfool}) with varying parameters to ensure \sys{}'s generalizability. The perturbation levels are selected such that the adversarial noise is undetectable by a human observer (Table~\ref{tab:attack_params} summarizes the pertinent attack parameters).} 
 
 
 \edit{There is a trade-off between the false positive and the true positive detection rates that can be controlled using the security parameter (see Section~\ref{framework}). The Area Under Curve (AUC) for a TP versus FP plot is a measure of accuracy for adversarial detection. A random decision has an AUC score of 0.5 while an ideal detector will have an AUC score of 1. Figure~\ref{fig:auc} shows the AUC score obtained by \sys{} for different attack configurations where the adversary knows everything about the model but is not aware of the defenders. For a given number of defenders, the AUC score for MNIST is relatively higher compared to more complex benchmarks (e.g., CIFAR-10). This is consistent with our hypothesis since the unexplored sub-space is larger in higher-dimensional benchmarks. Note that using more defenders eventually increases the AUC score. }
 

 \begin{table}[ht!]
\centering
\caption{Attack parameters: For CarliniL2 attack~\cite{carlini2017towards}, ``C'' denotes the confidence, ``LR'' is the learning rate, ``steps'' is the number of binary search steps, and ``iterations'' stands for the maximum number of iterations. Superscripts ($m\rightarrow{\text{MNIST}}$, $s\rightarrow{\text{SVHN}}$, $c\rightarrow{\text{CIFAR-10}}$, $a\rightarrow{\text{all}}$) are used to indicate the benchmarks for which the parameters are used.}
\vspace{-0.7em}
\label{tab:attack_params}
 \resizebox{0.96\columnwidth}{!}{
\begin{tabular}{c||c}
 \textbf{Attack}            & \textbf{Attack Parameters}\\ \hline \hline
 \textbf{FGS}               & $\epsilon \in \{0.01^{a},0.05^{a},0.1^{m,c},0.2^{m}\}$ \\  \hline
 \textbf{Deepfool}          & $n_{iters} \in \{2^{a},5^{a},10^{a},20^{a},50^{a},100^{a}\}$ \\    \hline
 \textbf{BIM} & $\epsilon \in \{0.001^{a},0.002^{a}\}, n_{iters}\in\{5^{a},10^{a},20^{a},50^{m},100^{m}\}$\\ \hline
 \textbf{CarliniL2} & \begin{tabular}[c]{@{}c@{}}$C \in \{0^{a},10^{a},20^{s,c},30^{s,c},40^{s,c},50^{c},60^{c},70^{c}\}$\\ LR = $0.1^{a}$, steps = $10^{a}$, iterations = $500^{a}$\end{tabular} \\

                          
                         
                                                               
\end{tabular}}
\vspace{-1.5em}
\end{table}





\subsection{Performance Analysis}{\label{sec:implementation}}
We implement the customized defender modules on \textit{Xilinx Zynq-ZC702} and \textit{Xilinx UltraScale-VCU108} FPGA platforms.
All modules are synthesized using \textit{Xilinx Vivado v2017.2}. We integrate the synthesized modules into a system-level block diagram with required peripherals, such as the DRAM, using Vivado IP Integrator. The frequency is set to $150~MHz$ and power consumption is estimated using the synthesis tool. 
For comparison purposes, we evaluate \sys{} performance against a highly-optimized TensorFlow-based implementation on two low-power embedded boards: (i)~The \textit{Jetson TK1} development kit which contains an \textit{NVIDIA Kepler} GPU with 192 CUDA Cores as well as an \textit{ARM Cortex-A15} 4-core CPU. (ii)~ A more powerful \textit{Jetson TX2} board which is equipped with an \textit{NVIDIA Pascal} GPU with 256 cores and a 6-core \textit{ARM v8} CPU.



\begin{figure}[ht!]
\centering
\includegraphics[width=0.8\columnwidth]{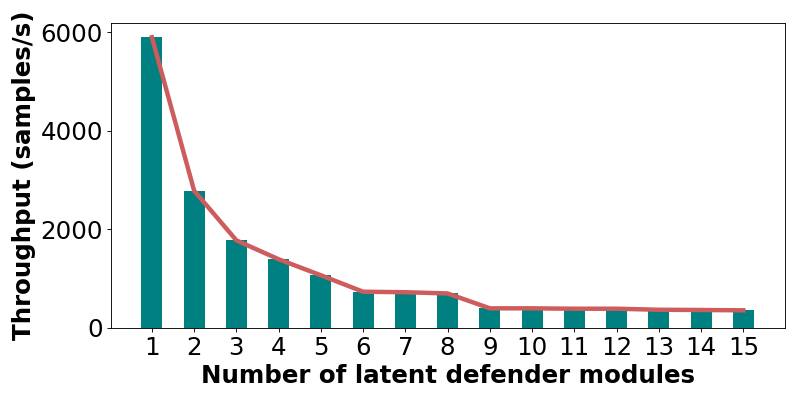}{\centering}
\caption{Throughput of \sys{} with samples from the MNIST dataset, implemented on the \textit{Xilinx Zync-ZC702} FPGA versus the number of instantiated defenders.}
\vspace{-0.5em}
  \label{fig:robust_throughput}
\end{figure}

\vspace{0.2em}
\noindent\textbf{Robustness and throughput trade-off.}
Increasing the number of checkpoints improves the reliability of model prediction in presence of adversarial attacks (Section~\ref{attack_analysis}) at the cost of reducing the effective throughput of the system. In applications with severe resource constraints, it is crucial to optimize system performance to ensure maximum immunity while adhering to the user-defined timing constraints. In scenarios with more flexible timing budget, the customization tool automatically allocates more instances of the defender modules while under strict timing constraints, the robustness is decreased in favor of the throughput. Figure~\ref{fig:robust_throughput} demonstrates the throughput versus the number of defender modules for MNIST benchmark on Zynq FPGA. The defender modules are located at the second-to-last layer of the victim DNN. Here the PCA kernel in the defender modules reduces the dimensionality to $10$.

Consider the SVHN benchmark, with the same throughput of 1400 samples per second, \sys{} implementation on UltraScale FPGA can run 8 defenders in parallel while the ARM v8 CPU can maintain the same throughput with only one defender. This directly translates to an improvement in the AUC score from 0.76 to 0.96.

\begin{figure}[ht!]
\centering
\includegraphics[width=0.49\columnwidth]{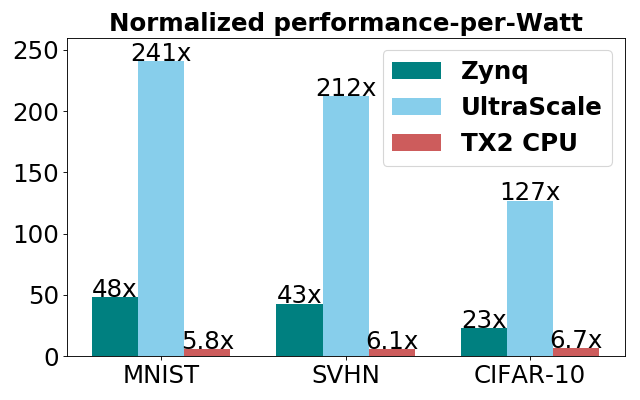}
\includegraphics[width=0.49\columnwidth]{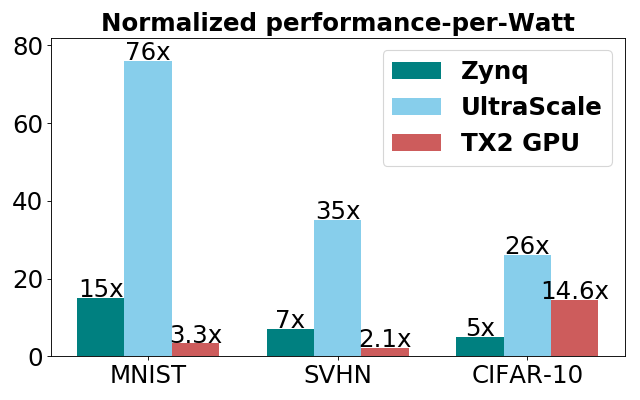}
\vspace{-0.5em}
\caption{Performance-per-Watt comparison with embedded CPU (left) and CPU-GPU (right) platforms. Reported values are normalized by the performance-per-Watt of \textit{Jetson TK1}.}\label{fig:throughput_GPU}
\vspace{-1em}
\end{figure}

\vspace{0.2em}
\noindent{\bf Throughput and energy analysis.}
To corroborate the efficiency of \sys{} framework, we also evaluate MRR performance on \textit{Jetson TK1} and \textit{Jetson TX2} boards operating in CPU-GPU and CPU-only modes. We define the performance-per-Watt measure as the throughput over the total power consumed by the system. This metric is an effective representation of the system performance since it integrates two influential factors for embedded system applications, namely the throughput and the power consumption. All evaluations in this section are performed with only one instance of the input and latent defenders. Figure~\ref{fig:throughput_GPU} (left) illustrates the performance-per-Watt for different hardware platforms. Numbers are normalized by the performance-per-Watt for the \textit{Jetson TK1} platform. As shown, \sys{} implementation on Zynq shows an average of $38\times$ improvement over the \textit{Jetson TK1} and $6.2\times$ improvement over the \textit{Jetson TX2} in the CPU mode. The more expensive UltraScale FPGA performs relatively better with an average improvement of $193\times$ and $31.7\times$ over the \textit{Jetson TK1} and \textit{Jetson TX2} boards, respectively.

The comparisons with GPU platforms are delineated in Figure~\ref{fig:throughput_GPU} (right). All values are normalized against the \textit{Jetson TK1} performance-per-Watt in the CPU-GPU mode. The evaluations show an average of $9\times$ and $45.7\times$ improvement over \textit{Jetson TK1} by the Zynq and UltraScale FPGAs, respectively. Comparisons with the \textit{Jetson TX2} demonstrate $2.74\times$ and $41.5\times$ improvement for the Zynq and UltraScale implementations. Note that the UltraScale performs noticeably better than the Zynq FPGA which emphasizes the effect of resource constraints on parallelism and the throughput.
\section{Related Work}
\edit{In response to the various adversarial attack methodologies proposed in the literature (e.g.,~\cite{goodfellow2014explaining, moosavi2016deepfool, carlini2017towards}), several research attempts have been made to design DL strategies that are more robust in the face of adversarial examples. The existing countermeasures can be classified into two categories: 
\textbf{(i)}~Supervised strategies which leverage the noise-corrupted inputs~\cite{ gu2014towards} and/or adversarial examples~\cite{ shaham2015understanding, goodfellow2014explaining, szegedy2013intriguing} during training of a DL model. These countermeasures are particularly tailored for specific perturbation patterns and can only partially evade adversarial samples generated by other attack scenarios (with different perturbation distributions) from being effective as shown in~\cite{gu2014towards}.
\textbf{(ii)}~Unsupervised approaches which aim to address adversarial attacks by smoothing out the gradient space (decision boundaries)~\cite{miyato2015distributional, carlini2017towards} or compressing the DL model by removing the nuisance variables~\cite{papernot2016distillation}. These set of works have been mainly remained oblivious to the data density in the latent space and are shown to be vulnerable to adaptive attacks where the adversary knows the defense mechanism~\cite{carlini2016defensive}. More recently,~\cite{meng2017magnet} proposes an unsupervised manifold projection method called MagNet to reform adversarial samples using auto-encoders. As shown in~\cite{carlini2017magnet}, manifold projection methods including MagNet are not robust to adversarial samples and can approximately increase the required distortion to generate adversarial sample by only 30\%. }

\sys{} methodology (called MRR) is an unsupervised approach that significantly improves the robustness of DL models against best-known adversarial attacks to date. Unlike \sys{}, no prior work has addressed resource efficiency or online performance of their defense algorithm.

\section{Conclusion}
This paper presents \sys{}, a novel end-to-end framework for online accelerated defense against adversarial samples in the context of deep learning. \edit{We introduce modular robust redundancy as a viable unsupervised countermeasure to significantly reduce the risk of integrity attacks.} To ensure applicability to various deep learning tasks and FPGA platforms, \sys{} provides an API that takes as input the high-level description of a deep neural network together with the specifications of the underlying hardware platform. Using a software-hardware-algorithm co-design approach, our automated customization tool optimizes the defense layout to maximize model reliability (safety) while complying with the hardware and/or user constraints. Our extensive evaluations corroborate the effectiveness and practicality of \sys{} framework.





{\bibliographystyle{IEEEtran}
\bibliography{ref}}

\begin{thebibliography}{10}
\providecommand{\url}[1]{#1}
\csname url@samestyle\endcsname
\providecommand{\newblock}{\relax}
\providecommand{\bibinfo}[2]{#2}
\providecommand{\BIBentrySTDinterwordspacing}{\spaceskip=0pt\relax}
\providecommand{\BIBentryALTinterwordstretchfactor}{4}
\providecommand{\BIBentryALTinterwordspacing}{\spaceskip=\fontdimen2\font plus
\BIBentryALTinterwordstretchfactor\fontdimen3\font minus
  \fontdimen4\font\relax}
\providecommand{\BIBforeignlanguage}[2]{{%
\expandafter\ifx\csname l@#1\endcsname\relax
\typeout{** WARNING: IEEEtran.bst: No hyphenation pattern has been}%
\typeout{** loaded for the language `#1'. Using the pattern for}%
\typeout{** the default language instead.}%
\else
\language=\csname l@#1\endcsname
\fi
#2}}
\providecommand{\BIBdecl}{\relax}
\BIBdecl

\bibitem{mcdaniel2016machine}
P.~McDaniel, N.~Papernot, and Z.~B. Celik, ``Machine learning in adversarial
  settings,'' \emph{IEEE Security \& Privacy}, vol.~14, no.~3, pp. 68--72,
  2016.

\bibitem{deng2014deep}
L.~Deng, D.~Yu \emph{et~al.}, ``Deep learning: methods and applications,''
  \emph{Foundations and Trends{\textregistered} in Signal Processing}, vol.~7,
  no. 3--4, pp. 197--387, 2014.

\bibitem{knorr2015paypal}
E.~Knorr, ``How paypal beats the bad guys with machine learning,'' 2015.

\bibitem{carlini2017towards}
N.~Carlini and D.~Wagner, ``Towards evaluating the robustness of neural
  networks,'' in \emph{IEEE Symposium on Security and Privacy (SP)}.\hskip 1em
  plus 0.5em minus 0.4em\relax IEEE, 2017, pp. 39--57.

\bibitem{goodfellow2014explaining}
I.~J. Goodfellow, J.~Shlens, and C.~Szegedy, ``Explaining and harnessing
  adversarial examples,'' \emph{arXiv preprint arXiv:1412.6572}, 2014.

\bibitem{kurakin2016adversarial}
A.~Kurakin, I.~Goodfellow, and S.~Bengio, ``Adversarial examples in the
  physical world,'' \emph{arXiv preprint arXiv:1607.02533}, 2016.

\bibitem{moosavi2016deepfool}
S.-M. Moosavi-Dezfooli, A.~Fawzi, and P.~Frossard, ``Deepfool: a simple and
  accurate method to fool deep neural networks,'' in \emph{IEEE Conference on
  Computer Vision and Pattern Recognition}, 2016, pp. 2574--2582.

\bibitem{meng2017magnet}
D.~Meng and H.~Chen, ``Magnet: a two-pronged defense against adversarial
  examples,'' in \emph{ACM SIGSAC Conference on Computer and Communications
  Security}.\hskip 1em plus 0.5em minus 0.4em\relax ACM, 2017, pp. 135--147.

\bibitem{zantedeschi2017efficient}
V.~Zantedeschi, M.-I. Nicolae, and A.~Rawat, ``Efficient defenses against
  adversarial attacks,'' in \emph{Proceedings of the 10th ACM Workshop on
  Artificial Intelligence and Security}.\hskip 1em plus 0.5em minus 0.4em\relax
  ACM, 2017, pp. 39--49.

\bibitem{shen2017ape}
S.~Shen, G.~Jin, K.~Gao, and Y.~Zhang, ``Ape-gan: Adversarial perturbation
  elimination with gan,'' \emph{ICLR Submission, available on OpenReview},
  2017.

\bibitem{carlini2017magnet}
N.~Carlini and D.~Wagner, ``Magnet and" efficient defenses against adversarial
  attacks" are not robust to adversarial examples,'' \emph{arXiv preprint
  arXiv:1711.08478}, 2017.

\bibitem{zhang2015optimizing}
C.~Zhang, P.~Li, G.~Sun, Y.~Guan, B.~Xiao, and J.~Cong, ``Optimizing fpga-based
  accelerator design for deep convolutional neural networks,'' in
  \emph{ACM/SIGDA International Symposium on Field-Programmable Gate
  Arrays}.\hskip 1em plus 0.5em minus 0.4em\relax ACM, 2015.

\bibitem{chen2014diannao}
T.~Chen, Z.~Du, N.~Sun, J.~Wang, C.~Wu, Y.~Chen, and O.~Temam, ``Diannao: A
  small-footprint high-throughput accelerator for ubiquitous
  machine-learning,'' \emph{ACM Sigplan Notices}, vol.~49, no.~4, pp. 269--284,
  2014.

\bibitem{sharma2016dnnweaver}
H.~Sharma, J.~Park, E.~Amaro, B.~Thwaites, P.~Kotha, A.~Gupta, J.~K. Kim,
  A.~Mishra, and H.~Esmaeilzadeh, ``Dnnweaver: From high-level deep network
  models to fpga acceleration,'' in \emph{The Workshop on Cognitive
  Architectures}, 2016.

\bibitem{samragh2017customizing}
M.~Samragh, M.~Ghasemzadeh, and F.~Koushanfar, ``Customizing neural networks
  for efficient fpga implementation,'' in \emph{Field-Programmable Custom
  Computing Machines (FCCM)}.\hskip 1em plus 0.5em minus 0.4em\relax IEEE,
  2017.

\bibitem{rouhani2017deep3}
B.~D. Rouhani, A.~Mirhoseini, and F.~Koushanfar, ``Deep3: Leveraging three
  levels of parallelism for efficient deep learning,'' in \emph{Proceedings of
  the 54th Annual Design Automation Conference 2017}.\hskip 1em plus 0.5em
  minus 0.4em\relax ACM, 2017, p.~61.

\bibitem{tropp2007signal}
J.~Tropp, A.~C. Gilbert \emph{et~al.}, ``Signal recovery from random
  measurements via orthogonal matching pursuit,'' \emph{IEEE Transactions on
  Information Theory}, vol.~53, no.~12, pp. 4655--4666, 2007.

\bibitem{diez1993parameter}
F.~J. Diez, ``Parameter adjustment in bayes networks. the generalized noisy
  or--gate,'' in \emph{Uncertainty in Artificial Intelligence, 1993}.\hskip 1em
  plus 0.5em minus 0.4em\relax Elsevier, 1993, pp. 99--105.

\bibitem{rouhani2015ssketch}
B.~D. Rouhani, E.~M. Songhori, A.~Mirhoseini, and F.~Koushanfar, ``Ssketch: An
  automated framework for streaming sketch-based analysis of big data on
  fpga,'' in \emph{Field-Programmable Custom Computing Machines (FCCM)}.\hskip
  1em plus 0.5em minus 0.4em\relax IEEE, 2015.

\bibitem{gu2014towards}
S.~Gu and L.~Rigazio, ``Towards deep neural network architectures robust to
  adversarial examples,'' \emph{arXiv preprint arXiv:1412.5068}, 2014.

\bibitem{shaham2015understanding}
U.~Shaham, Y.~Yamada, and S.~Negahban, ``Understanding adversarial training:
  Increasing local stability of neural nets through robust optimization,''
  \emph{arXiv preprint arXiv:1511.05432}, 2015.

\bibitem{szegedy2013intriguing}
C.~Szegedy, W.~Zaremba, I.~Sutskever, J.~Bruna, D.~Erhan, I.~Goodfellow, and
  R.~Fergus, ``Intriguing properties of neural networks,'' \emph{arXiv preprint
  arXiv:1312.6199}, 2013.

\bibitem{miyato2015distributional}
T.~Miyato, S.-i. Maeda, M.~Koyama, K.~Nakae, and S.~Ishii, ``Distributional
  smoothing with virtual adversarial training,'' \emph{arXiv preprint
  arXiv:1507.00677}, 2015.

\bibitem{papernot2016distillation}
N.~Papernot, P.~McDaniel, X.~Wu, S.~Jha, and A.~Swami, ``Distillation as a
  defense to adversarial perturbations against deep neural networks,'' pp.
  582--597, 2016.

\bibitem{carlini2016defensive}
N.~Carlini and D.~Wagner, ``Defensive distillation is not robust to adversarial
  examples,'' \emph{arXiv preprint}, 2016.

\end{thebibliography}

\end{document}